\documentclass[reprint,aps,prb,amsmath,amssymb]{revtex4-2}
\usepackage{xcolor}
\usepackage{graphicx}
\graphicspath{{figures/}}
\newcommand{\jj}{\mathrm{j}}
\newcommand{\dd}{\mathrm{d}}
\newcommand{\Real}{\operatorname{Re}}
\newcommand{\Imag}{\operatorname{Im}}
\newcommand{\PV}{\operatorname{P.V.}}
\newcommand{\artanh}{\operatorname{artanh}}
\newcommand{\bX}{\mathbf{X}}
\newcommand{\bk}{\mathbf{k}}
\newcommand{\bp}{\mathbf{p}}
\newcommand{\bs}{\mathbf{s}}
\newcommand{\bu}{\mathbf{u}}
\newcommand{\bv}{\mathbf{v}}
\newcommand{\bG}{\mathbf{G}}
\newcommand{\bM}{\mathbf{M}}
\newcommand{\aC}{a_\mathrm{CL}}
\newcommand{\vT}{V_\mathrm{0T}}
\newcommand{\vL}{V_\mathrm{0L}}
\newcommand{\kT}{k_\mathrm{0T}}
\newcommand{\IPb}{\mathrm{IP}}

\usepackage[colorlinks=true,allcolors=blue]{hyperref}

\begin{document}

\title{A unified closed-form theory for elastic waves in polycrystals of any crystal symmetry}

\author{Ming Huang}
\email{ming.huang@sussex.ac.uk}
\affiliation{Department of Engineering and Design, University of Sussex, Brighton BN1 9QT, UK}
\affiliation{Department of Mechanical Engineering, Imperial College London, London SW7 2AZ, UK}

\author{Stanislav I. Rokhlin}
\affiliation{Department of Materials Science and Engineering, The Ohio State University, Columbus, OH 43221, USA}

\author{Michael J. S. Lowe}
\affiliation{Department of Mechanical Engineering, Imperial College London, London SW7 2AZ, UK}

\begin{abstract}
The coherent elastic wave in random polycrystals is described by the first-order smoothing approximation, whose mass operator remains an integral evaluated only numerically. Here we derive its closed form for equiaxed, untextured polycrystals of any crystal symmetry with an exponential correlation function, giving an explicit dispersion equation at all frequencies. Unlike the integral form, which fails beyond one inverse correlation radius from the real axis, the closed form holds deep in the complex wavenumber plane, where heavily attenuated roots lie. It is compact, reproduces the dispersion equations of Stanke and Kino for cubic grains, extends them to any symmetry and contains the Born and far-field approximations as special cases. It reveals the structure of the roots and their partitions of the coherent field: three roots per polarization, one from the incident wave and one from each of the two scattering channels, without and with mode conversion. It also yields closed-form limits of their velocity, attenuation and partition in the Rayleigh, stochastic and geometric regimes, which show why a single coherent wave has been observed at long wavelengths, whereas more than one can be observed at short wavelengths. These results underpin further theory and quantitative characterization of polycrystalline microstructure.
\end{abstract}

\maketitle

\section{Introduction}\label{sec:intro}

Random polycrystals, from metals and ceramics to rocks, scatter elastic waves at the boundaries between grains of different crystallographic orientation. This scattering gives the coherent wave, i.e.\ the field averaged over the random grain structure, a complex wavenumber. Its real part gives the phase velocity, and its imaginary part gives the attenuation coefficient. Ultrasonic nondestructive evaluation uses this wavenumber to characterize grain size and texture and to determine the smallest detectable defect~\cite{Papadakis1965,Thompson2008,Rokhlin2021}, and seismology uses it to study the heterogeneous crust, mantle and inner core of the Earth~\cite{Sato2012}. Both rely on a theory that predicts the complex wavenumber from the microstructure.

The standard theory starts from the Dyson equation for the mean Green's function, which collects all multiple scattering in a mass operator~\cite{Frisch1968,Rytov1989,Weaver1990}. The first-order smoothing approximation (FOSA), or Bourret approximation, keeps the mass operator to second order in the fluctuations of the elastic constants~\cite{Bourret1962,Karal1964,Stanke1984,Weaver1990,Roy2025}, and the roots of the resulting dispersion equation give the complex wavenumbers of the coherent wave. The FOSA mass operator is a three-dimensional integral over the scattered wave vectors, which has been evaluated in several ways. Stanke and Kino~\cite{Stanke1984} integrated it in real space for cubic grains with an exponential correlation function and obtained explicit but lengthy dispersion equations. The Born approximation evaluates it at the wavenumber of the reference medium, which gives the attenuation~\cite{Weaver1990} and the phase velocity~\cite{Huang2020jasa} explicitly. Calvet and Margerin~\cite{Calvet2012} computed the spectral function on the real wavenumber axis. The far-field approximation (FFA) of Rokhlin \emph{et al.}~\cite{Rokhlin2015} simplifies the mass operator into a rational function of the wavenumber, so that the dispersion equation can be analyzed explicitly~\cite{Huang2021}. Most recently, Sha \emph{et al.}~\cite{Sha2020} and Huang \emph{et al.}~\cite{Huang2022a,Huang2022b} reduced the mass operator to a one-dimensional integral, valid for grains of any symmetry, which they evaluated numerically to find the roots at complex wavenumbers.

The FOSA dispersion equation has recently been found to have three roots for each polarization~\cite{Rokhlin2015,Huang2021,Huang2026jmps}, and grain-scale numerical simulations have resolved two of them as separate coherent waves once the wavelength becomes comparable to the grain size~\cite{Huang2026prl,Huang2026jmps}. The physical meaning of these roots has so far been explained with the FFA~\cite{Huang2026jmps}, because the FOSA mass operator contains a principal-value integral, which makes its dispersion equation transcendental. The FFA makes the structure of the roots explicit, but its two simplifications, the omission of the near-field part of the Green's function and the choice of a representative inner product, leave open which features belong to the FOSA itself. A closed form of the FOSA mass operator is therefore needed, both to reveal the number, physical meaning and observability of its roots without further approximation and to give an explicit dispersion equation for grains of any symmetry. Such an equation is also needed in practice, to compute the velocity and attenuation quickly, to interpret experiments and simulations, and to infer the microstructure from measurements.

In this paper, we derive the FOSA mass operator in closed form for equiaxed, untextured polycrystals of any crystal symmetry with an exponential correlation function. The closed form is compact, extends the dispersion equations of Stanke and Kino from cubic grains to any crystal symmetry and contains the Born approximation and the FFA as special cases. Unlike the integral form used so far, it remains valid in the whole complex wavenumber plane. It makes the structure of the FOSA dispersion equation explicit. We use it to count the roots, to give each a physical meaning and to estimate which can be observed as separate coherent waves. We also derive closed-form expressions for their velocity, attenuation and partition, i.e.\ their contribution to the coherent wave, from the Rayleigh to the geometric regime. Together, these results explain within the FOSA itself why one coherent wave has been observed at long wavelengths and more than one at short wavelengths. We focus on the theory; how well the FOSA describes polycrystals has been examined against grain-scale finite-element simulations elsewhere~\cite{Ryzy2018,Bai2020,Huang2022a,Huang2022b,Dorval2025,Huang2026jmps}.

The paper is organized as follows. Section~\ref{sec:theory} summarizes the FOSA, derives the closed form of its mass operator, gives the physical meaning of its terms and shows that only the closed form remains valid at all complex wavenumbers. Section~\ref{sec:analytic} explains the roots of the dispersion equation, derives their velocity, attenuation and partition in the Rayleigh, stochastic and geometric regimes, and relates the closed form to the Born and far-field approximations. Section~\ref{sec:conclusion} summarizes and concludes.

\section{Theory}\label{sec:theory}

We first introduce the FOSA and the integral form of its mass operator in Sec.~\ref{sec:fosa}. We then derive the closed form and give the physical meaning of its terms in Sec.~\ref{sec:closedform}, and show that only the closed form remains valid everywhere at complex wavenumbers in Sec.~\ref{sec:strip}.

\subsection{First-order smoothing approximation and its integral form}\label{sec:fosa}

We consider an untextured polycrystal formed by densely packed, fully bonded, equiaxed single-phase grains. The mass density $\rho$ is constant throughout, and the elastic tensor $c_{ijkl}(\bX)$ varies with position $\bX$ because the crystallographic orientation changes from grain to grain. Fields vary in time as $\exp(-\jj\omega t)$, with $\omega=2\pi f$ the angular frequency and $\jj^2=-1$. Elastic wave propagation in the polycrystal is governed by the equation of motion for the Green's function $G_{k\alpha}(\bX,\bX';\omega)$, which is the displacement in direction $k$ at $\bX$ due to a unit point force in direction $\alpha$ at $\bX'$~\cite{Karal1964,Weaver1990,Rokhlin2015},
\begin{equation}\label{eq:motion}
\begin{split}
  \left[\frac{\partial}{\partial X_i}\,\frac{c_{ijkl}(\bX)}{\rho}\,\frac{\partial}{\partial X_l}+\omega^2\delta_{jk}\right]&G_{k\alpha}(\bX,\bX')\\
  &=\delta_{j\alpha}\,\delta^3(\bX-\bX'),
\end{split}
\end{equation}
where $\delta_{j\alpha}$ is the Kronecker delta, $\delta^3$ the three-dimensional delta function, and repeated indices are summed. The Cartesian indices $k$ and $\alpha$ are unrelated to the wavenumber $k$ and the attenuation coefficient $\alpha$ used below. The elastic tensor is written as $c_{ijkl}(\bX)=C_{ijkl}+\delta c_{ijkl}(\bX)$, where the Voigt average $C_{ijkl}=\langle c_{ijkl}\rangle$ defines a homogeneous reference medium and $\delta c_{ijkl}(\bX)$ is a fluctuation of zero mean. Here $\langle\cdot\rangle$ represents the ensemble average over the grain orientations. Because the polycrystal is untextured, the reference medium is isotropic, with longitudinal and transverse velocities $\vL$ and $\vT$.

The coherent wave is described by the mean Green's function $\langle\bG\rangle$, which obeys the Dyson equation~\cite{Frisch1968,Weaver1990}. Because the polycrystal is statistically homogeneous, the Dyson equation becomes algebraic in the spatial Fourier domain,
\begin{equation}\label{eq:dyson}
  \langle\bG(\bk)\rangle=\left[\bG^0(\bk)^{-1}-\bM(\bk)\right]^{-1},
\end{equation}
where $\bk=k\bp$ is the wave vector with the unit propagation direction $\bp$. The bare Green's function $\bG^0$ is the Green's function of the reference medium. The mass operator $\bM$ describes the effect of all the scattering by the grains on the mean field. The Dyson equation is exact, but its mass operator is unknown. The FOSA approximates the mass operator by the leading term of its expansion in powers of the fluctuations, $\bM\simeq\langle\delta\mathcal{L}\,\bG^0\,\delta\mathcal{L}\rangle$, where $(\delta\mathcal{L})_{jk}=\partial_i\,(\delta c_{ijkl}/\rho)\,\partial_l$ is the fluctuating part of the operator in Eq.~\eqref{eq:motion}~\cite{Bourret1962,Frisch1968}. Each factor $\delta\mathcal{L}$ represents a scattering event, and $\bG^0$ carries the wave from one event to the other. The FOSA mass operator is therefore second order in the fluctuations, but solving Eq.~\eqref{eq:dyson} makes the mean Green's function include any number of such double-scattering events in succession.

The only statistical property of the polycrystal that enters the FOSA is the two-point covariance of the fluctuations. For orientations that are uncorrelated between grains and independent of the grain shapes, it factorizes as~\cite{Stanke1984,Weaver1990}
\begin{equation}\label{eq:covariance}
  \langle\delta c_{ijkl}(\bX)\,\delta c_{\alpha\beta\gamma\delta}(\bX')\rangle
  =\langle\delta c_{ijkl}\,\delta c_{\alpha\beta\gamma\delta}\rangle\,w(r),
\end{equation}
where $\langle\delta c_{ijkl}\,\delta c_{\alpha\beta\gamma\delta}\rangle$ is the covariance of the single-crystal constants and $w(r)$ is the two-point correlation function of the grains. The function $w(r)$ is the probability that two points lie in the same grain, and because the grains are equiaxed, it depends only on the distance $r=|\bX-\bX'|$ between the points. We take the exponential $w(r)=\exp(-r/a)$, with the correlation radius $a$. The spectral density of $w$, i.e.\ its three-dimensional Fourier transform as a function of the wave vector $\mathbf{q}$ with $q=|\mathbf{q}|$, is
\begin{equation}\label{eq:W}
  W(\mathbf{q})=\frac{1}{(2\pi)^3}\int w(r)\,e^{-\jj\mathbf{q}\cdot\mathbf{r}}\,\dd^3r
  =\frac{a^3}{\pi^2\left(1+a^2q^2\right)^2}.
\end{equation}

The bare Green's function of the isotropic reference medium is a sum over three plane waves, one longitudinal and two transverse ($M=\mathrm{L},\mathrm{T1},\mathrm{T2}$), with unit polarizations $\bu^M$ and velocities $V_{0M}$~\cite{Turner1999,Rokhlin2015},
\begin{equation}\label{eq:G0}
  \bG^0(\bk)=\sum_M\frac{\bu^M\bu^M}{\omega^2-k^2V_{0M}^2},
\end{equation}
where $V_{0\mathrm{L}}=\vL$ and $V_{0\mathrm{T1}}=V_{0\mathrm{T2}}=\vT$. Because the polycrystal is statistically isotropic, the mass operator has the same structure, $\bM=\sum_Mm_M\bu^M\bu^M$, where $m_M=\bu^M\cdot\bM\cdot\bu^M$ is the mass operator projected onto the polarization $\bu^M$. Inserting both into Eq.~\eqref{eq:dyson} gives the mean Green's function as $\langle\bG\rangle=\sum_Mg_M\bu^M\bu^M$, whose component $g_M$ for polarization $M$ satisfies
\begin{equation}\label{eq:dispersion}
  \frac{1}{g_M(k,\omega)}=D_M(k,\omega)=\omega^2-k^2V_{0M}^2-m_M(k,\omega),
\end{equation}
where $D_M$ is the dispersion function. Because $\langle\bG\rangle$ does not couple different polarizations, an incident wave of polarization $M$ excites a coherent field of the same polarization, which is described by $g_M$ alone. The coherent waves are the poles of $g_M$, i.e.\ the roots $k$ of the dispersion equation
\begin{equation}\label{eq:dispeq}
  D_M(k,\omega)=\omega^2-k^2V_{0M}^2-m_M(k,\omega)=0.
\end{equation}
Each root with positive real and imaginary parts is a forward-propagating coherent wave, with phase velocity $V=\omega/\Real k$ and attenuation coefficient $\alpha=\Imag k$. For root~$i$, we write $k_i$, $V_i$ and $\alpha_i$.

Between the two scattering events in $\bM$, the wave is carried by $\bG^0$, which is the sum over the three waves of Eq.~\eqref{eq:G0}. The projected mass operator is therefore a sum over the waves $N$ into which the incident wave $M$ is scattered, $m_M=m_{M\to\mathrm{L}}+m_{M\to\mathrm{T}}$. We call each term a scattering channel: the channel $N=M$ is the own channel of the incident wave (e.g.\ T$\to$T), and the channel $N\neq M$ is its cross channel (e.g.\ T$\to$L), which describes mode conversion. The two transverse waves form one channel because they share the velocity $\vT$. Each channel is~\cite{Weaver1990,Rokhlin2015,Sha2020}
\begin{equation}\label{eq:channel}
\begin{split}
  m_{M\to N}(k,\omega)=\frac{k^2}{\rho^2}\int&\frac{(k^S)^2\,\IPb_{M\to N}(\bp,\bs)}{\omega^2-(k^S)^2V_{0N}^2}\\
  &\times W(\bk-k^S\bs)\,\dd^3k^S ,
\end{split}
\end{equation}
where $\bk^S=k^S\bs$ is the wave vector of the scattered wave with unit polarization $\bv^N$, and the inner product
\begin{equation}\label{eq:IPdef}
  \IPb_{M\to N}(\bp,\bs)=\langle\delta c_{ijkl}\,\delta c_{\alpha\beta\gamma\delta}\rangle\,
  u^M_\beta u^M_k\,p_\alpha p_l\,v^N_\gamma v^N_j\,s_\delta s_i,
\end{equation}
combines the covariance with the polarizations ($\bu^M$ and $\bv^N$) and the directions ($\bp$ and $\bs$) of the incident and scattered waves. For $M\to\mathrm{T}$, the two scattered transverse polarizations are summed, and for $\mathrm{T}\to N$, the two incident ones are averaged, which corresponds to the factor of 2 for transverse waves in our earlier work~\cite{Huang2020jasa,Huang2022a}. The factors $k^2$ and $(k^S)^2$ in Eq.~\eqref{eq:channel} come from the gradients in $\delta\mathcal{L}$ acting on the incident and scattered waves. The integrand is singular on the sphere $k^S=k_{0N}\equiv\omega/V_{0N}$, where the scattered wave is a propagating wave of the reference medium, with velocity $V_{0N}$ and no attenuation. This singularity is treated by adding an infinitesimal positive imaginary part to the frequency, $\omega\to\omega+\jj0^+$, which moves the singularity off the real axis of $k^S$ so that the scattered wave travels outward from the scatterer, as causality requires.

In spherical coordinates with the polar axis along $\bp$, the variables of Eq.~\eqref{eq:channel} are $k^S$, the scattering angle $\theta$ between $\bp$ and $\bs$, and the azimuth $\varphi$ of $\bs$ about $\bp$. The spectral density depends only on $k^S$ and $\theta$, so the integration over $\varphi$ replaces the inner product by its average over $\varphi$~\cite{Rokhlin2015,Sha2020},
\begin{equation}\label{eq:IP}
  \IPb_{M\to N}(\theta)=A_{MN}+B_{MN}\cos^2\theta+C_{MN}\cos^4\theta ,
\end{equation}
where the coefficients are quadratic invariants of the single-crystal elastic constants and are available for any crystal symmetry~\cite{Sha2020}. The symmetry of the grains enters the theory only through these coefficients.

The integral over $\theta$ is then collected in the function
\begin{equation}\label{eq:h}
\begin{split}
  &h_{M\to N}(k,\xi)=\frac{2}{\pi a}\int_0^\pi\IPb_{M\to N}(\theta)\sin\theta\\
  &\qquad\times\left(k^2-2k\xi k_{0N}\cos\theta+\xi^2k_{0N}^2+a^{-2}\right)^{-2}\dd\theta,
\end{split}
\end{equation}
where $\xi=k^S/k_{0N}$ is the normalized scattered wavenumber. Equation~\eqref{eq:channel} thus reduces to a single integral over $\xi$~\cite{Sha2020,Huang2022a,Huang2026jmps},
\begin{equation}\label{eq:integral}
\begin{split}
  m_{M\to N}=\frac{k^2k_{0N}^3}{\rho^2V_{0N}^2}\bigg[&\PV\!\int_0^\infty\frac{\xi^4\,h_{M\to N}(k,\xi)}{1-\xi^2}\,\dd\xi\\
  &-\jj\frac{\pi}{2}\,h_{M\to N}(k,1)\bigg],
\end{split}
\end{equation}
where the first term in the brackets is a principal-value integral and the second is a residue term. Both come from the singularity of the integrand at $\xi=1$, i.e.\ at $k^S=k_{0N}$. The principal-value integral excludes an infinitesimal interval around $\xi=1$, and the residue term is the contribution of that interval, whose sign is fixed by causality. Existing methods~\cite{Sha2020,Huang2022a} evaluate the dispersion function at complex $k$ by substituting complex $k$ directly into Eq.~\eqref{eq:integral}, with the integral over $\theta$ in Eq.~\eqref{eq:h} in closed form~\cite{Weaver1990} and that over $\xi$ numerically.

The two-point correlation function of a real or simulated microstructure is better represented by a sum of exponentials fitted to its measured values~\cite{Sha2020},
\begin{equation}\label{eq:tpc}
  w(r)=\sum_n w_n\exp(-r/a_n),\qquad \aC=\sum_n w_na_n,
\end{equation}
with $\sum_nw_n=1$ and $\aC$ the correlation length. Because the mass operator is linear in $w$, its value for such a fit is the $w_n$-weighted sum over the terms of the fit, each with its own radius $a_n$.

\subsection{Closed form of the mass operator}\label{sec:closedform}

Because the spectral density of Eq.~\eqref{eq:W} is a rational function (a ratio of two polynomials) of the wavenumber, the integrals in Eq.~\eqref{eq:integral} can be evaluated exactly, which gives the mass operator in closed form. We evaluate the integral over $\xi$ by contour integration and then make the integral over the scattering angle rational by a change of variable, as detailed in Appendix~\ref{app:derivation}. The result is expressed through the complex wavenumber $\kappa_N=k_{0N}+\jj/a$, which is the wavenumber of wave $N$ in the reference medium with an added attenuation coefficient of $1/a$. It describes a wave that travels at the velocity $V_{0N}$ and decays over one correlation radius. Each scattering channel has the closed form
\begin{equation}\label{eq:closed}
\begin{split}
  m_{M\to N}=-\frac{k^2}{\rho^2V_{0N}^2}\bigg[&\left(A_{MN}+B_{MN}+C_{MN}\right)\frac{k_{0N}^2}{k^2-\kappa_N^2}\\
  &+A_{MN}+B_{MN}F_1+C_{MN}F_2\bigg],
\end{split}
\end{equation}
where
\begin{align}
  F_1&=1+\frac{\jj(k_{0N}-\jj/a)}{ak^2}-\frac{\jj\left(k^2+|\kappa_N|^2\right)}{ak^3}\artanh\frac{k}{\kappa_N},
  \label{eq:F1}\\
  F_2&=1+\frac{\jj}{2ak_{0N}}+\frac{\jj\left(5a^2k_{0N}^2-6\jj ak_{0N}+3\right)}{3a^3k_{0N}k^2}\notag\\
  &\quad+\frac{\jj\left[\left(1+a^2k_{0N}^2\right)^2-\jj ak_{0N}\left(2+a^2k_{0N}^2\right)\right]}{2a^5k_{0N}k^4}\notag\\
  &\quad-\frac{\jj\left(k^2+|\kappa_N|^2\right)^3}{2ak_{0N}^2k^5}\artanh\frac{k}{\kappa_N}\notag\\
  &\quad+\frac{\left(1+a^2k^2\right)^3}{2a^7k_{0N}^2k^5}\arctan(ka) ,
  \label{eq:F2}
\end{align}
where $\artanh$ is the inverse hyperbolic tangent. Because $\artanh$ and $\arctan$ are multivalued at complex arguments, Appendix~\ref{app:singular} specifies which of their values to use. Equation~\eqref{eq:closed} holds for grains of any crystal symmetry through the coefficients of Eq.~\eqref{eq:IP}, and, for the multi-term correlation function of Eq.~\eqref{eq:tpc}, it applies to each term with $a$ replaced by $a_n$.

Each channel in Eq.~\eqref{eq:closed} is the sum of three terms, which we call the pole, constant and angular terms. The pole term, $-k^2\IPb_{M\to N}(0)\,k_{0N}^2/[\rho^2V_{0N}^2(k^2-\kappa_N^2)]$, is a simple pole at $k=\kappa_N$, whose strength is the inner product in the forward direction, $\IPb_{M\to N}(0)=A_{MN}+B_{MN}+C_{MN}$. The constant term, $-k^2A_{MN}/(\rho^2V_{0N}^2)$, contributes to the mass operator at all frequencies with the same coefficient $A_{MN}$. The angular term, $-k^2(B_{MN}F_1+C_{MN}F_2)/(\rho^2V_{0N}^2)$, carries the angular dependence of scattering and contains the functions $\artanh$ and $\arctan$. The physical meaning of the three terms follows from splitting the scattered wave in Eq.~\eqref{eq:channel} exactly into two parts (Appendix~\ref{app:derivation}): a local part, in which the wave is scattered twice at the same point without propagating in between, and a propagating part, which carries the scattered wave from one scattering point to another. The constant term is the local part for the angle-independent part $A_{MN}$ of the inner product. The pole term comes from the propagating part alone: in real space, the outgoing scattered wave is multiplied by the correlation function $\exp(-r/a)$, so it becomes a wave of wavenumber $\kappa_N$, which places the pole at $\kappa_N$. The angular term collects the remaining contributions of the angle-dependent part of the inner product, from both the local and the propagating parts.

For cubic crystals, Eq.~\eqref{eq:closed}, inserted into Eq.~\eqref{eq:dispeq}, reproduces the dispersion equations of Stanke and Kino~\cite{Stanke1984} once the misprints in their transverse equation are corrected~\cite{Rokhlin2015}. For other crystal symmetries, it extends these equations, with only the coefficients $A_{MN}$, $B_{MN}$ and $C_{MN}$ changed. For all crystal symmetries, Eq.~\eqref{eq:closed} gives the same mass operator as the integral form in Eq.~\eqref{eq:integral} for $|\Imag k|<1/a$, as we confirmed numerically. The closed form has two advantages over the integral form. It needs no numerical integration, and it remains valid for $|\Imag k|\ge1/a$, where the integral form fails (Sec.~\ref{sec:strip}). Deep in the Rayleigh regime ($ka\lesssim0.01$), however, the closed-form terms in $1/k^2$ to $1/k^5$ of $F_1$ and $F_2$ become large and nearly cancel, so that their much smaller sum is lost to rounding error. We remove this cancellation by expanding $\artanh$ and $\arctan$ in their Taylor series, whose leading terms cancel those terms analytically.

\subsection{Validity of the integral and closed forms}\label{sec:strip}

Because the roots of the dispersion equation are complex, we now evaluate the range of validity of the integral and closed forms in representing the mass operator in the complex wavenumber plane. As an example, we use transverse waves in iron, a material with cubic crystals ($c_{11}=219.2$, $c_{12}=136.8$, $c_{44}=109.2$~GPa, $\rho=7860$~kg/m$^3$), whose Voigt velocities are $\vL=5900$~m/s and $\vT=3230$~m/s.

The integral form of Eq.~\eqref{eq:integral} is valid for complex $k$ only when its integrand stays finite on the whole path of integration. However, the spectral density enters Eq.~\eqref{eq:h} as the inverse square of $k^2-2k\xi k_{0N}\cos\theta+\xi^2k_{0N}^2+a^{-2}$. Whenever $|\Imag k|\ge1/a$, this expression vanishes at some point of the path, where the integrand becomes infinite (Appendix~\ref{app:singular}). The integral form therefore gives the mass operator correctly only in the strip
\begin{equation}\label{eq:strip}
  |\Imag k|<1/a,
\end{equation}
or $|\Imag k|<1/\max_na_n$ for the multi-term correlation function of Eq.~\eqref{eq:tpc}. The closed form contains no integral and has no such limit. It is valid in the whole complex plane, apart from three branch points in the upper half-plane, $\kappa_\mathrm{L}$, $\kappa_\mathrm{T}$ and $\jj/a$, and the vertical branch cuts that run from them to infinity, which do not affect the roots (Appendix~\ref{app:singular}).

\begin{figure}[!ht]
\centering
\includegraphics[width=\columnwidth]{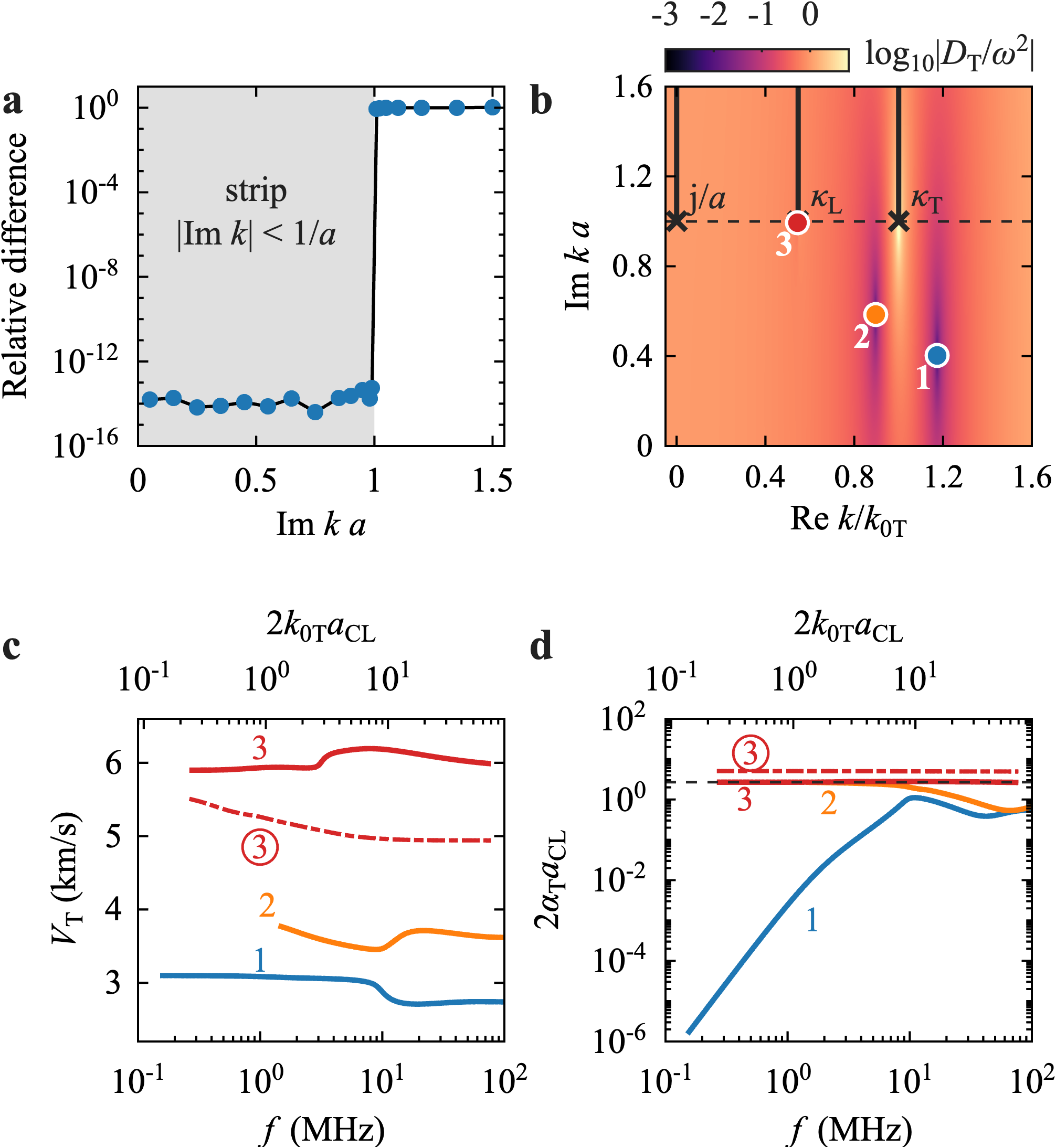}
\caption{Validity of the integral and closed forms for transverse waves in iron. (a)~Relative difference between the integral form, Eq.~\eqref{eq:integral}, and the closed form, Eq.~\eqref{eq:closed}, versus $\Imag k\,a$, at 30~MHz and $\Real k=0.95\kT$. The correlation function is a single exponential with $a=0.2309$~mm. Shaded: the strip of Eq.~\eqref{eq:strip}. (b)~$\log_{10}|D_\mathrm{T}/\omega^2|$ from the closed form in the complex wavenumber plane, for the same frequency and correlation function. Circles: roots~1--3. Crosses and vertical lines: branch points $\kappa_\mathrm{L}$, $\kappa_\mathrm{T}$ and $\jj/a$, and their branch cuts. Dashed line: edge of the strip. (c)~Phase velocity and (d)~normalized attenuation $2\alpha\aC$ of roots~1--3 versus frequency, for the eight-term fit of Ref.~\onlinecite{Huang2026jmps}. Solid: closed form. Dash-dotted, circled~3: root~3 from the integral form. Dashed: edge of the strip, $\Imag k=1/a_\mathrm{max}$. Top axes: normalized frequency $2\kT\aC$.}\label{fig:complex}
\end{figure}

Figure~\ref{fig:complex}(a) tests the two forms at 30~MHz for a single exponential correlation function with $a=0.2309$~mm. As the wavenumber moves into the complex plane at $\Real k=0.95\kT$, they agree to $5\times10^{-12}\%$ up to $\Imag k\,a=0.99$, but they differ by 90\% at 1.01 and by about 100\% beyond. The same failure occurs for longitudinal waves and for other materials, because the edge of the strip is set by the correlation radius alone.

The failure of the integral form matters because the heavily attenuated roots lie near the edge of the strip. Figure~\ref{fig:complex}(b) shows the dispersion function from the closed form in the complex wavenumber plane. It has exactly three roots (Appendix~\ref{app:singular}), numbered as in Sec.~\ref{sec:structure}: root~1 is the incident wave modified by scattering, and roots~2 and 3 start from the poles $\kappa_\mathrm{T}$ and $\kappa_\mathrm{L}$ of the own and cross channels. Roots~1 and 2 lie well inside the strip, whereas root~3 lies just inside its edge, at $\Imag k\,a=0.994$, near the point $\kappa_\mathrm{L}$. For a multi-term correlation function, the strip is narrower, and root~3 can lie outside it. An example is the eight-term fit to the correlation function of simulated microstructures~\cite{Huang2026jmps}, with the correlation length $\aC=0.2309$~mm. Roots~1 and 2 computed from the two forms agree to $10^{-8}\%$. However, the third root found with the integral form lies outside the strip and is therefore spurious, as shown in Fig.~\ref{fig:complex}(c,~d). The closed form places root~3 near $\kappa_\mathrm{L}$, as for the single exponential.

\section{Roots of the dispersion equation and their limits}\label{sec:analytic}

In this section, we use the closed form to explain the roots of the dispersion equation. Section~\ref{sec:structure} shows that the dispersion equation has three roots, gives their physical meaning and defines their partitions. Sections~\ref{sec:rayleigh}, \ref{sec:stochastic} and \ref{sec:geometric} derive closed-form expressions for the velocity, attenuation and partition of the roots in the Rayleigh, stochastic and geometric regimes, and compare them with earlier results. Section~\ref{sec:approx} shows that the Born and far-field approximations are special cases of the full theory.

\subsection{Number, meaning and partition of the roots}\label{sec:structure}

\begin{figure}[!ht]
\centering
\includegraphics[width=\columnwidth]{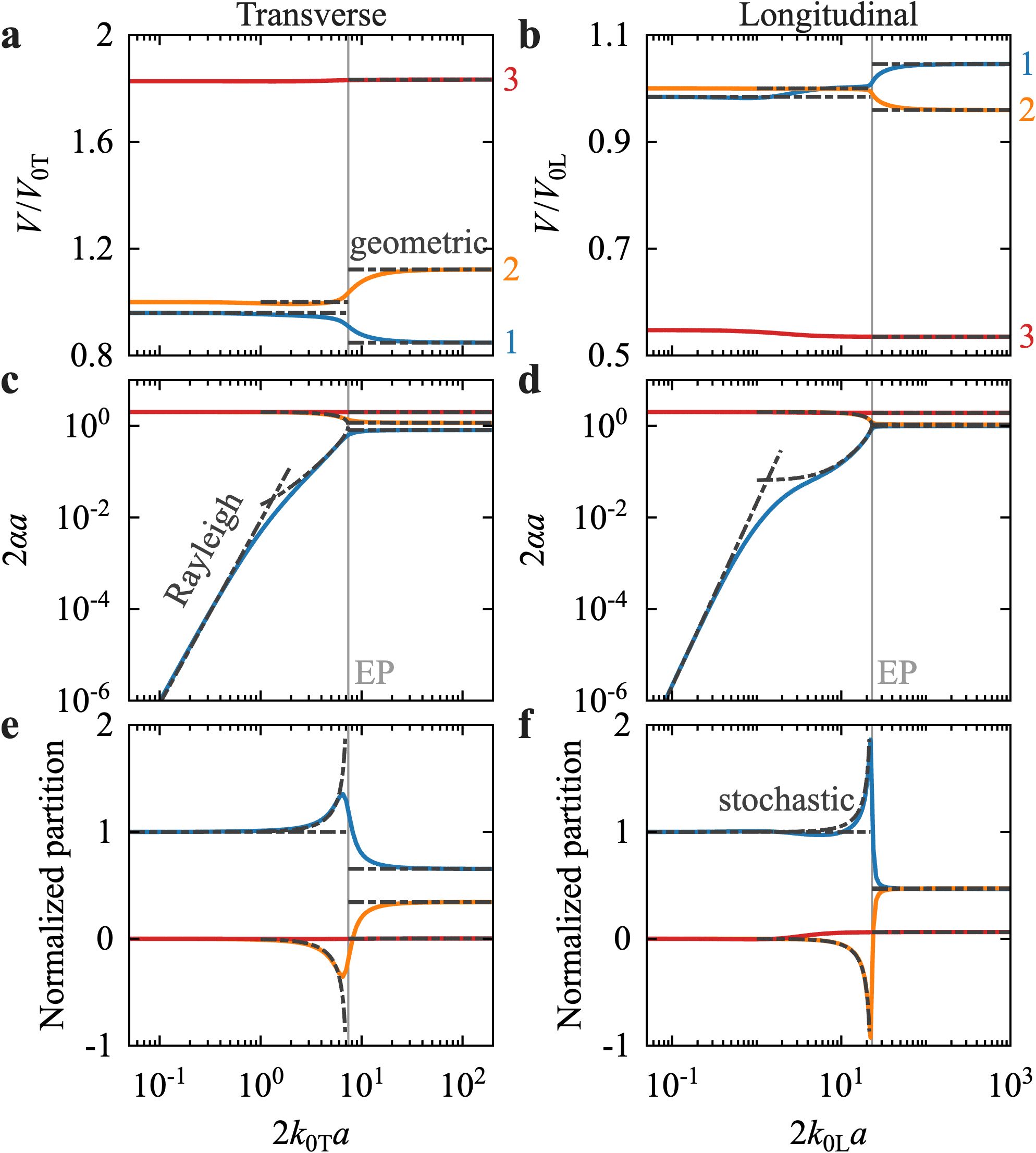}
\caption{Roots~1--3 of the dispersion equation for transverse (left) and longitudinal (right) waves in iron versus the normalized frequency $2k_{0M}a$, for a single exponential correlation function. (a,~b)~Phase velocity normalized by $V_{0M}$. (c,~d)~Normalized attenuation $2\alpha a$. (e,~f)~Normalized partition $\bar\eta$. Solid: full theory, with the roots numbered at the right of (a) and (b). Dash-dotted: Rayleigh, stochastic and geometric limits, labeled in the panels. The stochastic limits are Eqs.~\eqref{eq:stochroots}--\eqref{eq:stochpart}, with the constant from the cross channel added to the attenuation of root~1. Gray vertical line: exceptional point (EP), Eq.~\eqref{eq:EP}.}\label{fig:regimes}
\end{figure}

The dispersion equation has three roots, which can be inferred from the pole terms of Eq.~\eqref{eq:closed} as the scattering weakens. We use this weak-scattering limit only to show the number of roots and their physical meaning. It is not a further approximation of the theory: all the roots in this paper, including those in Fig.~\ref{fig:regimes}, are computed from the closed form without further approximation, which we call the full theory. In this weak-scattering limit, all the terms of the mass operator become small, except the pole terms near $k=\kappa_N$, which remain large there however weak the scattering, because their denominators vanish at $k=\kappa_N$. We therefore keep only the pole terms. With the forward scattering factors $Q_{M\to N}=\IPb_{M\to N}(0)/(4\rho^2V_{0M}^2V_{0N}^2)$, the mass operator becomes
\begin{equation}\label{eq:weakm}
  m_M\simeq-4V_{0M}^2k^2\sum_{N=\mathrm{L},\mathrm{T}}\frac{Q_{M\to N}\,k_{0N}^2}{k^2-\kappa_N^2},
\end{equation}
and the dispersion equation, Eq.~\eqref{eq:dispeq}, divided by $V_{0M}^2$, becomes
\begin{equation}\label{eq:weak}
  k_{0M}^2-k^2+4k^2\sum_{N=\mathrm{L},\mathrm{T}}\frac{Q_{M\to N}\,k_{0N}^2}{k^2-\kappa_N^2}=0.
\end{equation}
Multiplied by $(k^2-\kappa_\mathrm{L}^2)(k^2-\kappa_\mathrm{T}^2)$, this is a cubic equation in $k^2$. Each of its three roots in $k^2$ gives one physical root $k$, with positive real and imaginary parts. As $Q_{M\to N}\to0$, these three roots tend to $k_{0M}$, $\kappa_\mathrm{L}$ and $\kappa_\mathrm{T}$. One root therefore lies near the bare wavenumber $k_{0M}$, and one lies near each pole $\kappa_N$, which gives three roots for each polarization. The full dispersion equation has these three roots and only three, as the root count of Appendix~\ref{app:singular}, based on the argument principle, confirms at every frequency examined. This agrees with our earlier analysis based on the FFA~\cite{Huang2026jmps}. A multi-term correlation function adds a pole $k_{0N}+\jj/a_n$ for each term, and with it a further root near that pole. These further roots come from the fit rather than from the physics, and they are very strongly attenuated, with $\Imag k\simeq1/a_n$, so we do not consider them.

The roots have a direct physical meaning. The root near $k_{0M}$ is the effective wave of earlier theories, i.e.\ the incident wave modified by scattering. The root near $\kappa_N$ is the field scattered into channel $N$. It travels at the velocity $V_{0N}$ of the scattered wave and stays correlated with its source only over about one correlation radius, because $\Imag\kappa_N=1/a$. In the own channel ($N=M$), this field has the polarization of the incident wave, and in the cross channel ($N\neq M$), it is produced by mode conversion. We number the roots as in Ref.~\onlinecite{Huang2026jmps} and refer to them by these numbers throughout. Root~1 is the effective wave, which is observed at low frequency; root~2 is the second wave, which can be observed at a higher frequency; and root~3 is the remaining one. In iron, root~2 corresponds to the pole of the own channel and root~3 to the pole of the cross channel.

Whether a root can be observed is determined by its peak in the spectral function $-\Imag g_M(k)$ at real $k$, which describes how the coherent field is distributed over wavenumbers~\cite{Huang2026jmps}. The peak must be high enough and separated from the other peaks (Sec.~\ref{sec:geometric}). Near root $i$, the mean Green's function behaves as $g_M\simeq R_i/(k^2-k_i^2)$, where the residue is $R_i=2k_i/(\partial D_M/\partial k)$ at $k=k_i$. The closed form gives $\partial D_M/\partial k$ analytically. Each root therefore contributes to $g_M$ as a single wave of wavenumber $k_i$, whose amplitude relative to the bare wave of Eq.~\eqref{eq:G0}, $-V_{0M}^{-2}/(k^2-k_{0M}^2)$, is $-V_{0M}^2R_i$. For real $k$ near $\Real k_i$, $k^2-k_i^2\simeq2\Real k_i\,(k-\Real k_i-\jj\alpha_i)$, with $\alpha_i=\Imag k_i$, so this wave forms a peak centered at $\Real k_i$, with the half-width $\alpha_i$ and the height $H_i/V_{0M}^2$, where
\begin{equation}\label{eq:H}
  H_i=\frac{\eta_i}{2\alpha_i\Real k_i},
\end{equation}
and
\begin{equation}\label{eq:eta}
  \eta_i=-V_{0M}^2\Real R_i
\end{equation}
is the partition of root $i$, i.e.\ the real part of its relative amplitude, or the share of the coherent field that it carries. It equals 1 for the single wave of the reference medium. The imaginary part of $R_i$ only makes the peak asymmetric. A root therefore has a high peak only if it carries a sizable share of the field and is weakly attenuated; otherwise, it forms a low, broad peak that is hidden by the others. We also use the normalized partition $\bar\eta_i=\eta_i/\sum_j\eta_j$, so that the normalized partitions of the three roots add to 1.

Figure~\ref{fig:regimes} shows the velocity, attenuation and normalized partition of the three roots for transverse and longitudinal waves in iron, for a single exponential correlation function. The solid lines are from the full theory. The other lines are the limits derived in Secs.~\ref{sec:rayleigh}--\ref{sec:geometric}.

\subsection{Rayleigh regime}\label{sec:rayleigh}

In the low-frequency Rayleigh regime, the wavelength is much longer than the grain size ($k,k_{0N}\ll1/a$). We therefore expand the closed form, Eq.~\eqref{eq:closed}, in powers of $ka$ and $k_{0N}a$. The zeroth-order term is real, so it gives the velocity of root~1 but no attenuation. The attenuation comes from the leading imaginary part, which is of third order.

At zeroth order, the pole term vanishes, and $F_1\to\frac13$ and $F_2\to\frac15$. Because the grains are much smaller than the wavelength, the spectral density of Eq.~\eqref{eq:W} is nearly constant, $W\simeq a^3/\pi^2$, so the wave is scattered in all directions, and the inner product enters only through its average over all scattering directions. With the angle-averaged scattering factors $Q^*_{M\to N}=(A_{MN}+B_{MN}/3+C_{MN}/5)/(4\rho^2V_{0M}^2V_{0N}^2)$, as in our earlier work~\cite{Huang2022a}, Eq.~\eqref{eq:closed} becomes
\begin{equation}\label{eq:static}
  m_{M\to N}\to-4V_{0M}^2k^2\,Q^*_{M\to N}.
\end{equation}
The dispersion equation, Eq.~\eqref{eq:dispeq}, then becomes $\omega^2-k^2V_{0M}^2(1-4\sum_NQ^*_{M\to N})=0$, whose root gives the phase velocity of root~1 at zero frequency, which we call the quasi-static velocity,
\begin{equation}\label{eq:Vs}
  V_M=V_{0M}\Big(1-4\sum_NQ^*_{M\to N}\Big)^{1/2}.
\end{equation}
In iron, it lies 4.0\% below $\vT$ for transverse waves and 1.6\% below $\vL$ for longitudinal waves (Fig.~\ref{fig:regimes}(a,~b)). It does not depend on the grain size, because it comes entirely from the constant and angular terms of Eq.~\eqref{eq:closed}, whose zeroth-order values are independent of $a$. To first order in $Q^*_{M\to N}$, Eq.~\eqref{eq:Vs} reduces to two earlier results: the static correction $V_{0M}(1-2\sum_NQ^*_{M\to N})$ used with the FFA~\cite{Rokhlin2015}, and the Rayleigh velocity $V_{0M}/(1+2\sum_NQ^*_{M\to N})$ derived from the Born approximation~\cite{Huang2020jasa}. For transverse waves in iron, Eq.~\eqref{eq:Vs} differs from them by 0.08\% and 0.23\% of $\vT$, respectively.

At third order, the leading imaginary part of Eq.~\eqref{eq:closed} is
\begin{equation}\label{eq:rayleighIm}
  \Imag m_{M\to N}\simeq-8V_{0M}^2k^2(k_{0N}a)^3\,Q^*_{M\to N},
\end{equation}
which contains the same factors $Q^*_{M\to N}$ as Eq.~\eqref{eq:static}, so the velocity and the attenuation of root~1 are both set by the angle-averaged scattering. Because root~1 stays close to $k_{0M}$, its attenuation to first order in the scattering, obtained by expanding Eq.~\eqref{eq:dispeq} about $k=k_{0M}$, is $\alpha_1=-\Imag m_M(k_{0M},\omega)/(2k_{0M}V_{0M}^2)$, and Eq.~\eqref{eq:rayleighIm} gives
\begin{equation}\label{eq:rayleigh}
  \alpha_1=4a^3k_{0M}\sum_NQ^*_{M\to N}\,k_{0N}^3\propto\omega^4,
\end{equation}
which is proportional to the effective grain volume, $\int w(r)\,\dd^3r=8\pi a^3$, or $8\pi\sum_nw_na_n^3$ for a multi-term correlation function~\cite{Huang2022a}. This Rayleigh asymptote is identical to that derived earlier from the Born approximation~\cite{Huang2021jasa}. Equations~\eqref{eq:Vs} and \eqref{eq:rayleigh} are also the low-frequency limits of the equations of Stanke and Kino for cubic grains, and here we extend the velocity limit to any crystal symmetry.

For iron at 10~kHz, Eq.~\eqref{eq:rayleigh} agrees with the attenuation of the full theory to 0.01\% for both polarizations. Figure~\ref{fig:regimes}(c,~d) shows that it holds up to about $2k_{0M}a\simeq1$ for transverse waves, and only up to a lower frequency for longitudinal waves, whose attenuation is dominated by conversion into the slower transverse waves. The transverse waves produced by this conversion have a wavelength 1.8 times shorter than that of the incident longitudinal wave, so they leave the Rayleigh regime first.

With $D_M=\omega^2-k^2V_M^2$, the derivative is $\partial D_M/\partial k=-2kV_M^2$, and Eq.~\eqref{eq:eta} gives the partition $\eta_1=(V_{0M}/V_M)^2$. The partitions of roots~2 and 3 vanish, because the pole terms that produce these roots vanish as $(k_{0N}a)^2$. Hence $\bar\eta_1\to1$ (Fig.~\ref{fig:regimes}(e,~f)), as found numerically in Ref.~\onlinecite{Huang2026jmps}.

These limits explain why a single coherent wave is observed in the Rayleigh regime. Root~1 propagates at nearly $V_M$, with an attenuation that is small and grows as $\omega^4$, and it carries the whole coherent field. Roots~2 and 3 stay at their poles with $\Imag k\simeq1/a$ ($2\alpha a\simeq2$), so they decay within about one correlation radius, carry almost none of the field and form no visible peak.

\subsection{Stochastic regime}\label{sec:stochastic}

In the stochastic regime, the wavelength $2\pi/k_{0M}$ is comparable to the grain size, so $k_{0M}a$ is already several times larger than 1 ($k_{0M}a\sim2\pi$). The scattering is still weak enough for root~1 to stay close to $k_{0M}$. Near $k=k_{0M}$, the pole term of the own channel is then larger than all the other terms of the mass operator by a factor of order $k_{0M}a$. This is because $\kappa_M$ lies only $1/a$ from $k_{0M}$, so that the denominator $k^2-\kappa_M^2\simeq2k_{0M}(k-\kappa_M)$ is much smaller than $k_{0M}^2$ there. This term describes the forward scattering of the incident wave into its own channel, so the inner product enters only through its forward value, i.e.\ through $Q_{M\to M}$. We keep only this term of Eq.~\eqref{eq:closed} in the dispersion equation, Eq.~\eqref{eq:dispeq}, and write $\omega^2-k^2V_{0M}^2\simeq-2k_{0M}V_{0M}^2(k-k_{0M})$. The dispersion equation then reduces to
\begin{equation}\label{eq:twolevel}
  (k-k_{0M})(k-\kappa_M)=Q_{M\to M}\,k_{0M}^2,
\end{equation}
whose two roots are
\begin{equation}\label{eq:stochroots}
  k=k_{0M}+\frac{\jj}{2a}\mp\jj\sqrt{\frac{1}{4a^2}-Q_{M\to M}k_{0M}^2}.
\end{equation}
The upper sign gives root~1. The lower sign gives root~2, near $\kappa_M$. Both roots have the real part $k_{0M}$, so both travel at $V_{0M}$ in this approximation. Their attenuations are
\begin{equation}\label{eq:stochatt}
  2\alpha_{1,2}a=1\mp\left[1-4Q_{M\to M}(k_{0M}a)^2\right]^{1/2},
\end{equation}
and their partitions follow from the residues of Eq.~\eqref{eq:twolevel},
\begin{equation}\label{eq:stochpart}
  \eta_{1,2}=\frac12\left[1\pm\left(1-4Q_{M\to M}(k_{0M}a)^2\right)^{-1/2}\right],
\end{equation}
whose sum is 1. Scattering into the cross channel adds a constant to $2\alpha_1a$, $k_{0M}\Gamma_N(k_{0M})/\allowbreak(\rho^2V_{0M}^2V_{0N}^2k_{0N})$ with $N\neq M$, where $\Gamma_N$ is given by Eq.~\eqref{eq:Gamma} in Sec.~\ref{sec:geometric}. Root~3, which lies near the pole of the cross channel, is not described by Eq.~\eqref{eq:twolevel}. It stays near that pole, with $2\alpha_3a\simeq2$ and a small partition, so it decays within about one correlation radius and cannot be observed in this regime. These velocities, attenuations and partitions are the stochastic limits shown in Fig.~\ref{fig:regimes}.

The two roots of Eq.~\eqref{eq:stochroots} coincide when the square root vanishes, i.e.\ at
\begin{equation}\label{eq:EP}
  2k_{0M}a=1/\sqrt{Q_{M\to M}}.
\end{equation}
Such a point, at which two roots coincide, is called an exceptional point. There, both roots have $2\alpha a=1$, and their partitions become large and of opposite sign, as the residues of two nearly coincident roots do. The roots coincide only in the approximation of Eq.~\eqref{eq:stochroots}. In the full theory, however, the other terms of the mass operator keep the two roots apart, so they never coincide; instead, they get closest to each other near the exceptional point, and their partitions stay finite (Fig.~\ref{fig:regimes}(e,~f)). Above the exceptional point, the term $\mp\jj\sqrt{\cdot}$ in Eq.~\eqref{eq:stochroots} becomes real. The two roots then have the same attenuation, $2\alpha a=1$, and the same partition, $1/2$, but different velocities, approaching $V_{0M}(1\mp\sqrt{Q_{M\to M}})$ at high frequency. They become roots~1 and 2 of the geometric regime (Sec.~\ref{sec:geometric}). The exceptional point thus marks the end of the stochastic regime and the beginning of the geometric regime.

Well below the exceptional point of Eq.~\eqref{eq:EP}, i.e.\ for $2k_{0M}a\sqrt{Q_{M\to M}}\ll1$, the limits of Eqs.~\eqref{eq:stochatt} and \eqref{eq:stochpart} simplify. The attenuations become
\begin{equation}\label{eq:stoch}
\begin{aligned}
  2\alpha_1a&=2Q_{M\to M}(k_{0M}a)^2,\\
  2\alpha_2a&=2-2Q_{M\to M}(k_{0M}a)^2.
\end{aligned}
\end{equation}
The first is identical to the stochastic asymptote we derived earlier from the Born approximation by keeping only the forward scattering of the incident wave~\cite{Huang2021jasa}. It is proportional to the correlation length, $\int_0^\infty w(r)\,\dd r=a$, or $\aC$ for a multi-term correlation function. In addition, root~2 loses exactly the attenuation that root~1 gains, so that $2\alpha_1a+2\alpha_2a=2$. The partitions become
\begin{equation}\label{eq:stochpartsimp}
  \eta_1\simeq1+Q_{M\to M}(k_{0M}a)^2,\qquad\eta_2\simeq-Q_{M\to M}(k_{0M}a)^2.
\end{equation}
Root~2 therefore has a small negative partition, which grows in magnitude with frequency, and root~1 carries slightly more than the whole field, so that the two partitions still add to 1. A negative partition does not make the spectral function negative. Instead, root~2 makes the peak of root~1 asymmetric and does not form a peak of its own~\cite{Huang2026jmps}. These limits thus explain the negative partition of root~2 found numerically in Ref.~\onlinecite{Huang2026jmps}.

For transverse waves in iron ($Q_{\mathrm{T}\to\mathrm{T}}=0.018$), the exceptional point is at $2k_{0\mathrm{T}}a=7.4$, close to where root~2 moves away from its pole and to the onset of the second wave in simulations ($2k_{0\mathrm{T}}\aC\simeq9$)~\cite{Huang2026jmps}. The range between the end of the Rayleigh regime, near $2k_{0\mathrm{T}}a=1$, and the exceptional point is therefore narrow. In Fig.~\ref{fig:regimes}(a), root~2 travels close to $\vT$, as Eq.~\eqref{eq:stochroots} predicts, but root~1 stays several percent below $\vT$ and transitions directly from its quasi-static velocity toward its geometric value. In Fig.~\ref{fig:regimes}(c), Eq.~\eqref{eq:stochatt} follows the attenuations of roots~1 and 2 at the upper end of this range, whereas the simplified form of Eq.~\eqref{eq:stoch} matches root~1 only near $2k_{0\mathrm{T}}a=3$. In Fig.~\ref{fig:regimes}(e), Eq.~\eqref{eq:stochpart} follows the partitions up to near the exceptional point, where it diverges.

For longitudinal waves in iron ($Q_{\mathrm{L}\to\mathrm{L}}=0.0019$), the exceptional point is at $2k_{0\mathrm{L}}a=23$, well above the end of the Rayleigh regime, and a stochastic regime forms between them. In Fig.~\ref{fig:regimes}(b), roots~1 and 2 both travel within 0.2\% of $\vL$, as Eq.~\eqref{eq:stochroots} predicts. In Fig.~\ref{fig:regimes}(d), the attenuation of root~1 follows Eq.~\eqref{eq:stochatt} with the constant from the cross channel to within 10\% from $2k_{0\mathrm{L}}a=8$ up to the exceptional point. Below $2k_{0\mathrm{L}}a=8$, the constant (0.064) exceeds the contribution of the own channel, because longitudinal waves are attenuated mainly by conversion into transverse waves. These converted waves leave the Rayleigh regime first (Sec.~\ref{sec:rayleigh}), so the contribution of the cross channel to the velocity and attenuation of root~1 levels off at a lower frequency than that of the own channel. This causes the step in the velocity of root~1 in Fig.~\ref{fig:regimes}(b) and the shoulder in its attenuation in Fig.~\ref{fig:regimes}(d), which is the hump noted by Stanke and Kino. Because of this shoulder, the attenuation does not follow the $\omega^2$ law of Eq.~\eqref{eq:stoch}, and Stanke and Kino concluded that longitudinal waves in iron have no stochastic region~\cite{Stanke1984}. Equation~\eqref{eq:stochatt}, with the constant from the cross channel, still describes this regime. In Fig.~\ref{fig:regimes}(f), root~2 has the small negative partition of Eq.~\eqref{eq:stochpart}, to within 5\%.

In general, a stochastic regime exists only if the exceptional point lies well above $2k_{0M}a\sim1$, i.e.\ for $Q_{M\to M}\ll1$, and it is wider for weaker scattering into the own channel. This agrees with Stanke and Kino, who placed the upper end of the stochastic region at $2k_{0M}a$ equal to the inverse of their inhomogeneity parameter and found that this region shortens as the grain anisotropy grows~\cite{Stanke1984}. For longitudinal waves, their inhomogeneity parameter equals $\sqrt{Q_{\mathrm{L}\to\mathrm{L}}}$, so Eq.~\eqref{eq:EP} reproduces their boundary between the stochastic and geometric regimes.

\subsection{Geometric regime}\label{sec:geometric}

In the geometric regime, the wavelength is much shorter than the grain size ($k_{0N}a\to\infty$), and the frequency lies far above the exceptional point. For $k$ of the order of $k_{0\mathrm{L}}$ and $k_{0\mathrm{T}}$, $F_1$ and $F_2$ tend to 1 with corrections of order $1/(k_{0N}a)$, and $\kappa_N\to k_{0N}$, because $1/a\ll k_{0N}$. The mass operator becomes
\begin{equation}\label{eq:geomM}
  m_{M\to N}\to-\frac{k^2\,\IPb_{M\to N}(0)}{\rho^2V_{0N}^2}\,\frac{k^2}{k^2-k_{0N}^2},
\end{equation}
in which the pole, constant and angular terms all have the forward inner product as their coefficient, because a wave much shorter than the grains is scattered mainly forward. The dispersion function, Eq.~\eqref{eq:dispersion}, is then real for real $k$. With $k=\omega/V$ and the forward scattering factors $Q_{M\to N}$ of Sec.~\ref{sec:structure}, the dispersion equation becomes
\begin{equation}\label{eq:geomcubic}
  V^2=V_{0M}^2-\sum_{N=\mathrm{L},\mathrm{T}}\frac{4Q_{M\to N}V_{0M}^2V_{0N}^2}{V_{0N}^2-V^2}.
\end{equation}
Multiplied by $(\vL^2-V^2)(\vT^2-V^2)$, this is a cubic equation in $V^2$. Its three solutions $V_i$ are the phase velocities that roots~1--3 approach at high frequency; they are constants independent of frequency and grain size, set by the forward scattering factors alone. In iron, they are 2739, 3623 and 5920~m/s (roots~1--3) for transverse waves, and 6168, 5661 and 3157~m/s for longitudinal waves. When the incident wave scatters mainly into its own channel, roots~1 and 2 have the squared velocities $V_{0M}^2(1\mp2\sqrt{Q_{M\to M}})$, which correspond to local stiffnesses one standard deviation below and above the mean (Appendix~\ref{app:pair}). Physically, the additional roots originate from the crystal nature of the grains. At high frequency, a grain much larger than the wavelength supports three waves of different polarizations, one quasi-longitudinal and two quasi-transverse, whose speeds depend on the orientation of the grain. These grain-scale waves couple across the grain boundaries and contribute to the coherent waves~\cite{Huang2026jmps}.

The attenuation of the roots follows from the next order in $1/(k_{0N}a)$. We expand Eq.~\eqref{eq:closed} to first order in $1/a$ at real $k$ away from the poles, using $\arctan(ka)\simeq\pi/2-1/(ka)$ and the values of $\artanh(k/\kappa_N)$ given in Appendix~\ref{app:singular}. At zeroth order, the roots $k_i=\omega/V_i$ are real. The first-order term moves them only slightly: its real part barely changes the velocities, and its imaginary part gives each root a small attenuation. Expanding $D_M$ about $k_i$ to first order gives the attenuation of root $i$,
\begin{equation}\label{eq:geomatt}
  2\alpha_ia=-\frac{2}{\partial D_M/\partial k}\sum_N\frac{k_i^2\,\Gamma_N(k_i)}{\rho^2V_{0N}^2k_{0N}},
\end{equation}
where $\partial D_M/\partial k$ is evaluated at $k_i$ from Eqs.~\eqref{eq:dispersion} and \eqref{eq:geomM}, and
\begin{equation}\label{eq:Gamma}
\begin{split}
  \Gamma_N(k)={}&\frac{2\IPb_{M\to N}(0)\,k_{0N}^4}{(k^2-k_{0N}^2)^2}\\
  &+B_{MN}\frac{k_{0N}\left[kk_{0N}-(k^2+k_{0N}^2)\Lambda\right]}{k^3}\\
  &+C_{MN}\frac{3k^4+10k^2k_{0N}^2+3k_{0N}^4}{6k^4}\\
  &-C_{MN}\frac{(k^2+k_{0N}^2)^3\Lambda}{2k_{0N}k^5},
\end{split}
\end{equation}
with $\Lambda=\Real\artanh(k/k_{0N})$. The first term of $\Gamma_N$ comes from the pole term, i.e.\ from the decay of the scattered wave over one correlation radius, and the terms in $B_{MN}$ and $C_{MN}$ come from the angular term. Like the velocities, the normalized attenuations $2\alpha_ia$ do not depend on frequency, but the attenuation coefficients $\alpha_i$ are inversely proportional to the grain size. For a multi-term correlation function, $1/a$ is replaced by $-w'(0)=\sum_nw_n/a_n$, the magnitude of the initial slope of the correlation function. Because $w(r)$ is the probability that two points a distance $r$ apart lie in the same grain, $-w'(0)$ is the number of grain boundaries crossed per unit length of a straight line, i.e.\ $1/a_\mathrm{MC}$, the inverse of the mean chord length $a_\mathrm{MC}$ of the grains (the mean linear intercept in metallography)~\cite{Stanke1986,Man2006,Sha2020}. The values of $2\alpha_ia$ therefore hold for any correlation function when $a$ is replaced by $a_\mathrm{MC}$. For transverse waves in iron, $2\alpha_1a=0.81$ and $2\alpha_2a=1.16$, so the slower root is the less attenuated.

Because the dispersion function is real in this limit, the residues are real, and the partitions of Sec.~\ref{sec:structure} are
\begin{equation}\label{eq:geomweight}
  \eta_i=-\frac{2k_iV_{0M}^2}{\partial D_M/\partial k},\qquad\sum_{i=1}^3\eta_i=\frac{1}{1-4\sum_NQ_{M\to N}},
\end{equation}
where the sum follows from $D_M\simeq-k^2V_{0M}^2(1-4\sum_NQ_{M\to N})$ at large $k$. For transverse waves in iron, the normalized partitions are 0.65, 0.34 and 0.003, close to the values 0.69, 0.31 and 0.004 computed numerically in Ref.~\onlinecite{Huang2026jmps} at 30~MHz with the fitted correlation function.

For root~1, an attenuation that does not depend on frequency and is inversely proportional to the grain size is the classical geometric behavior, found for cubic grains by Stanke and Kino~\cite{Stanke1984}. Their value for transverse waves in iron, 0.8 in our normalization, agrees with $2\alpha_1a=0.81$. Here we derive, for the first time, these limits in closed form for any crystal symmetry, together with those of roots~2 and 3, and show that the velocities do not depend on the grain size. Figure~\ref{fig:regimes} shows that the full theory approaches these limits at high frequency. For transverse waves in iron, Eqs.~\eqref{eq:geomatt} and \eqref{eq:geomweight} agree with the full theory to 0.01\% for $2k_{0M}a\ge300$.

A root is observed as a separate coherent wave, i.e.\ as a separate pulse in grain-scale numerical simulations~\cite{Huang2026prl,Huang2026jmps}, only if its peak in the spectral function (Sec.~\ref{sec:structure}) is both separated from the others and high enough. The geometric limits estimate both conditions.

\textit{Peak separation.} Two roots can be resolved when their peaks are separated by more than the sum of their half-widths, $|\Real(k_1-k_2)|>\Imag k_1+\Imag k_2$. With $\Real k_i=\omega/V_i$ from the geometric limits, this condition gives the frequency above which a second wave can be observed,
\begin{equation}\label{eq:onset}
  2k_{0M}a>\frac{2a(\alpha_1+\alpha_2)}{V_{0M}|1/V_1-1/V_2|}.
\end{equation}
For iron, it gives $2k_{0M}a>6.8$ for transverse and 24 for longitudinal waves, within 8\% and 4\% of the exceptional points of 7.4 and 23 from Eq.~\eqref{eq:EP}.

\textit{Peak height.} A separated root is seen only if its peak is high enough. All three roots have a real velocity and an attenuation of order $1/a$, so each of them can in principle be observed, and the height $H_i$ of its peak (Sec.~\ref{sec:structure}), which combines its partition and attenuation, decides whether it is. For transverse waves in iron, root~3 carries 0.3\% of the coherent field and has the largest attenuation (Fig.~\ref{fig:regimes}(c,~e)), so its peak is about 300 times lower than that of root~1. For longitudinal waves in iron, root~3 carries 6\% of the field (Fig.~\ref{fig:regimes}(f)), through the strong conversion into transverse waves, and its peak is about 30 times lower. The peak of root~3 thus rises as the cross channel strengthens. A third observable wave would therefore require a stronger cross channel, i.e.\ stronger mode conversion, than in iron, as in strongly anisotropic materials.

\subsection{Relation to the Born and far-field approximations}\label{sec:approx}

For equiaxed grains, the Born approximation and the FFA are special cases of the full closed-form theory, although both also apply to statistically elongated grains~\cite{Huang2020jasa,Rokhlin2015}. As in earlier studies, the Born approximation evaluates the mass operator at the bare wavenumber $k_{0M}$ and keeps only the first order in the scattering, so that the root of Eq.~\eqref{eq:dispeq} is $k\simeq k_{0M}-m_M(k_{0M},\omega)/(2k_{0M}V_{0M}^2)$~\cite{Weaver1990,Huang2020jasa}. It therefore describes root~1 alone. The real part of this root gives the phase velocity, $V_1=V_{0M}/[1-\Real m_M(k_{0M},\omega)/(2k_{0M}^2V_{0M}^2)]$, which we derived in this explicit form earlier, together with its Rayleigh asymptote (Sec.~\ref{sec:rayleigh})~\cite{Huang2020jasa}. The imaginary part gives the attenuation, $\alpha_1=-\Imag m_M(k_{0M},\omega)/(2k_{0M}V_{0M}^2)$. In Eq.~\eqref{eq:integral}, the velocity comes from the principal-value integral and the attenuation from the residue term. The attenuation reduces to the Rayleigh and stochastic asymptotes of Secs.~\ref{sec:rayleigh} and \ref{sec:stochastic}.

The FFA of Rokhlin \emph{et al.}~\cite{Rokhlin2015} replaces the bare Green's function by its far-field form, $\exp(\jj k_{0N}r)/r$, and neglects the near-field terms that dominate close to the scatterer. It therefore keeps only the propagating part of the mass operator. It also takes the inner product out of the integral at a representative value $\mathrm{Ip}_{M\to N}$, in the notation of Ref.~\onlinecite{Rokhlin2015}, which is the forward value for the own channel and the angular average for the cross channel. Its channel term is
\begin{equation}\label{eq:ffa}
  m^\mathrm{FFA}_{M\to N}=-\frac{k^2\,\mathrm{Ip}_{M\to N}}{\rho^2V_{0N}^2}\,\frac{k_{0N}^2}{k^2-\kappa_N^2},
\end{equation}
which is the pole term of Eq.~\eqref{eq:closed} with $\mathrm{Ip}_{M\to N}$ in place of the forward inner product. The dispersion equation of the FFA for equiaxed grains~\cite{Rokhlin2015} therefore has the same form as Eq.~\eqref{eq:weak}, with $\mathrm{Ip}_{M\to N}$ in place of $\IPb_{M\to N}(0)$ in $Q_{M\to N}$. The full closed-form theory thus confirms that the FFA retains the term of the mass operator that produces the roots, with its exact position $\kappa_N$. In the full closed-form theory, the strength of this pole is the forward inner product in every channel, which for the cross channel of transverse waves in iron is 19\% larger than the angular average. The constant and angular terms, which the FFA leaves out, give the quasi-static velocity, and the FFA includes it through its static correction (Sec.~\ref{sec:rayleigh}). In the geometric regime, the constant and angular terms shift the velocities of the FFA~\cite{Rokhlin2015,Huang2026jmps}, 2819, 3690 and 5917~m/s for transverse waves in iron, by up to 3\%, to those of Eq.~\eqref{eq:geomcubic}. They also separate the attenuations of roots~1 and 2, which the FFA gives as $2\alpha a=1/\sqrt{1+Q_{M\to M}}=0.99$~\cite{Huang2026jmps}, into 0.81 and 1.16 from Eq.~\eqref{eq:geomatt}.

\section{Summary and conclusions}\label{sec:conclusion}

We have derived the closed form of the mass operator in the dispersion equation of the first-order smoothing approximation for coherent elastic waves in equiaxed, untextured polycrystals of any crystal symmetry with an exponential correlation function. The closed form is compact, reproduces the dispersion equations of Stanke and Kino for cubic grains, extends them to any crystal symmetry and contains the Born and far-field approximations as special cases. Unlike the integral form used in existing methods, which holds only in the strip $|\Imag k|<1/a$, it is valid in the whole complex wavenumber plane, as required for the heavily attenuated roots. The closed form allowed us to show that the dispersion equation has three roots and to give their physical meaning and partitions, i.e.\ their contributions to the coherent wave. It also allowed us to derive closed-form limits of their velocities, attenuations and partitions in the Rayleigh, stochastic and geometric regimes.

The work leads to four findings.
\begin{enumerate}
 \item The incident wave $M$ scatters through two channels: its own channel, $M\to M$ (e.g.\ T$\to$T), and the cross channel, $M\to N$ with $N\neq M$ (e.g.\ T$\to$L), i.e.\ mode conversion. Each channel is the sum of a pole, a constant and an angular term. The pole terms give the dispersion equation three roots for each polarization, one for the incident wave and one for each of the two channels. A root can be observed as a separate coherent wave only if its peak in the wavenumber spectrum of the coherent field is separated from the others and high enough, which depends on both its partition and its attenuation.
 \item In the Rayleigh regime, for any crystal symmetry, the dispersion equation reduces to root~1, which carries the whole coherent field. The wave is scattered in all directions, so the velocity and attenuation of root~1 are set by the angle-averaged scattering factors. Its quasi-static velocity, obtained here in closed form, does not depend on the grain size, and its attenuation grows as the fourth power of frequency and is proportional to the effective grain volume.
 \item In the stochastic regime, the incident wave scatters mainly forward into its own channel, $M\to M$. As a result, the dispersion equation reduces to roots~1 and 2. Both travel at about the Voigt velocity. The attenuation of root~1 is proportional to the correlation length, as found earlier from the Born approximation, and root~2 has a small negative partition, so it forms no peak of its own but only makes the peak of root~1 asymmetric. The two roots get closest to each other, without coinciding, near an exceptional point. Above it, they separate in velocity, and the geometric regime begins. As the scattering into the own channel strengthens, the exceptional point moves down toward the end of the Rayleigh regime, so a stochastic regime exists only when this scattering is weak, as Stanke and Kino found for strongly anisotropic grains~\cite{Stanke1984}.
 \item In the geometric regime, the dispersion equation keeps all three roots. The wave is scattered mainly forward, and the velocities, attenuations and partitions of the roots have simple closed forms that do not depend on frequency, for any crystal symmetry. The velocities do not depend on the grain size, whereas the attenuations are inversely proportional to the mean chord length of the grains, which generalizes the classical geometric attenuation of root~1.
\end{enumerate}

The closed form replaces all numerical integration in the evaluation of the theory and makes the dispersion equation as cheap to solve as that of the far-field approximation while keeping all the terms of the mass operator. Its newly derived limits express the velocity, attenuation and partition of every coherent wave in terms of the single-crystal elastic constants and the grain size. They provide fast and transparent forward models for the ultrasonic characterization of polycrystals~\cite{Papadakis1965,Thompson2008,Rokhlin2021}. In particular, the velocity limits depend on the grain anisotropy but not on the grain size, whereas the attenuation limits depend on both: on the anisotropy through the scattering factors, and on the grain size through the effective grain volume, the correlation length and the mean chord length in the three regimes. Measurements of velocity and attenuation can therefore separate the effects of grain anisotropy and grain size and allow both to be characterized reliably.

\section*{Author declarations}

\subsection*{Conflict of interest}

The authors have no conflicts of interest to disclose.

\section*{Data availability}

The MATLAB code that implements the closed form and generates the results of this study is openly available in Zenodo at \url{https://doi.org/10.5281/zenodo.24366443}.

\appendix

\section{Closed form and its terms}\label{app:derivation}

\textit{Derivation of the closed form.} We evaluate the integral form of Eq.~\eqref{eq:integral} exactly, for a single exponential correlation function, in two steps. We first evaluate the integral over $\xi$ by contour integration and then make the remaining integral over the scattering angle rational by a change of variable. For both steps, we use $\hat q=k/k_{0N}$, $b=1/(k_{0N}a)$ and $\mu=\cos\theta$, and write the bracket in Eq.~\eqref{eq:h} as $k_{0N}^2\,\mathcal{D}(\xi,\mu)$, where
\begin{equation}\label{eq:Dxi}
\begin{gathered}
  \mathcal{D}(\xi,\mu)=\xi^2-2\hat q\mu\,\xi+\hat q^2+b^2=(\xi-\xi_+)(\xi-\xi_-),\\
  \xi_\pm=\hat q\mu\pm\jj\beta,\qquad \beta=\sqrt{\hat q^2+b^2-\hat q^2\mu^2}.
\end{gathered}
\end{equation}
First, we recombine the principal-value and residue terms of Eq.~\eqref{eq:integral} into a single integral over $\xi$ with the factor $1/(1-\xi^2+\jj0^+)$, whose imaginary part is fixed by causality. The integrand is unchanged by $(\xi,\mu)\to(-\xi,-\mu)$, so the integral over $\xi$ can be extended to the whole real line, which gives
\begin{equation}\label{eq:F}
\begin{aligned}
  m_{M\to N}&=\frac{k^2}{\pi\rho^2V_{0N}^2\,ak_{0N}}\int_{-1}^{1}\IPb_{M\to N}(\mu)\,\mathcal{F}(\mu)\,\dd\mu,\\
  \mathcal{F}(\mu)&=\int_{-\infty}^{\infty}\frac{\xi^4\,\dd\xi}{(1-\xi^2+\jj0^+)\,\mathcal{D}(\xi,\mu)^2}.
\end{aligned}
\end{equation}
For real $k$, the integrand of $\mathcal{F}$ is a rational function that decays as $\xi^{-2}$, so we evaluate the integral by contour integration, closing the path with a large semicircle in the upper half of the complex $\xi$ plane. The integral then equals $2\pi\jj$ times the sum of the residues at the two poles enclosed. These are the pole $\xi=1+\jj0^+$ of the bare Green's function, displaced off the real axis by causality, and the zero $\xi_+$ of $\mathcal{D}$, which is a double pole of the integrand because $\mathcal{D}$ appears squared. Hence
\begin{equation}\label{eq:residues}
\begin{aligned}
  \mathcal{F}(\mu)&=2\pi\jj\left[-\frac{1}{2\mathcal{D}(1,\mu)^2}+\mathcal{R}_+(\mu)\right],\\
  \mathcal{R}_+(\mu)&=\frac{\dd}{\dd\xi}\frac{\xi^4}{(1-\xi^2)(\xi-\xi_-)^2}\bigg|_{\xi=\xi_+}.
\end{aligned}
\end{equation}
The residue at $\xi=1$ reproduces the residue term of Eq.~\eqref{eq:integral} exactly, and the residue at $\xi_+$ replaces the principal-value integral.

Second, we make the remaining integral over $\mu$ an integral of a rational function by the substitution $\mu=(\tau/\hat q)\sin\psi$, with $\tau=\sqrt{\hat q^2+b^2}$, which gives $\beta=\tau\cos\psi$ and $\xi_\pm=\pm\jj \tau\,e^{\mp\jj\psi}$. In the variable $z=e^{-\jj\psi}$, the residue at $\xi_+$, $\mu$ and $\dd\mu$ become
\begin{equation}
\begin{gathered}
  \mathcal{R}_+=\frac{2\jj \tau\,z^5\left[(\tau^2+1)z^2+2\right]}{(z^2+1)^3(\tau^2z^2+1)^2},\\
  \mu=\frac{\tau\,(z^{-1}-z)}{2\jj \hat q},\qquad
  \dd\mu=\frac{\jj \tau\,(z^2+1)}{2\hat q\,z^2}\,\dd z .
\end{gathered}
\end{equation}
As $\mu$ runs from $-1$ to 1, the path runs along the unit circle from $z_1=(b+\jj \hat q)/\tau$ to $z_2=(b-\jj \hat q)/\tau$. The integrals $\int\mu^{2n}\mathcal{R}_+\,\dd\mu$ with $n=0$, 1 and 2, which multiply $A_{MN}$, $B_{MN}$ and $C_{MN}$, respectively, are integrals of rational functions of $z$, which give rational terms and the logarithms $\ln(\tau^2z^2+1)$ and $\ln z$. The first term in the brackets of Eq.~\eqref{eq:residues}, which is the residue at $\xi=1$, gives integrals of powers of $\mathcal{D}(1,\mu)=1+\hat q^2+b^2-2\hat q\mu$, which is linear in $\mu$, and hence rational terms and the logarithm $\ell=\ln\{[(1+\hat q)^2+b^2]/[(1-\hat q)^2+b^2]\}$. In physical variables, these logarithms combine into $\artanh(k/\kappa_N)$ and $\arctan(ka)$, which gives Eqs.~\eqref{eq:closed}--\eqref{eq:F2}.

\textit{Physical meaning of the three terms.} The three terms of Eq.~\eqref{eq:closed} follow from splitting the integrand of Eq.~\eqref{eq:channel} in two ways. First, because $\omega^2=k_{0N}^2V_{0N}^2$, the factor of the integrand that describes the scattered wave splits exactly into two parts,
\begin{equation}\label{eq:split}
  \frac{(k^S)^2}{\omega^2-(k^S)^2V_{0N}^2}=-\frac{1}{V_{0N}^2}+\frac{k_{0N}^2}{\omega^2-(k^S)^2V_{0N}^2}.
\end{equation}
The first part does not depend on $k^S$, so in real space it is a delta function at the scattering point, i.e.\ the local part. It exists only because the gradients in $\delta\mathcal{L}$ give the numerator $(k^S)^2$. The second part is $k_{0N}^2$ times the bare Green's function, which in real space is proportional to the outgoing wave $e^{\jj k_{0N}r}/r$ and is the propagating part. Second, the inner product splits into its angle-independent part $A_{MN}$ and its angle-dependent part $B_{MN}\cos^2\theta+C_{MN}\cos^4\theta$. Their combinations give the three terms.
\begingroup\setlength{\topsep}{0pt}\setlength{\partopsep}{0pt}\setlength{\leftmargini}{1.9em}
\begin{enumerate}\renewcommand{\labelenumi}{(\roman{enumi})}\setlength{\itemsep}{0pt}\setlength{\parsep}{0pt}\setlength{\labelsep}{0.4em}
\item Local part with $A_{MN}$: the constant term, because $\int W\,\dd^3q=w(0)=1$.
\item Propagating part with $A_{MN}$: a pole of strength $A_{MN}$ at $k=\kappa_N$, because in real space the outgoing wave is multiplied by $w(r)=\exp(-r/a)$, which gives $e^{\jj\kappa_Nr}/r$.
\item Both parts with the angle-dependent part: the angular term, and the remaining strength $B_{MN}+C_{MN}$ of the pole.
\end{enumerate}\endgroup
The local part does not depend on $k_{0N}$, so the pole term comes from the propagating part alone.

\section{Singularities and root count}\label{app:singular}

\textit{Singularities of the integral form.} The spectral density in Eq.~\eqref{eq:h} becomes infinite where $k^2-2k\xi k_{0N}\cos\theta+\xi^2k_{0N}^2+a^{-2}=0$, i.e.\ at
\begin{equation}\label{eq:sing}
  k=\xi k_{0N}\cos\theta\pm\jj\sqrt{\xi^2k_{0N}^2\sin^2\theta+1/a^2},
\end{equation}
whose imaginary part is at least $1/a$ in magnitude, with equality in the forward and backward directions. For every $k$ with $|\Imag k|\ge1/a$, some point $(\xi,\theta)$ of the path is therefore singular, which gives the strip of Eq.~\eqref{eq:strip}. For a multi-term correlation function, each term has its own strip, and the narrowest, $|\Imag k|<1/\max_na_n$, applies. We evaluate the weighted sum of Eq.~\eqref{eq:closed} over the eight terms of the fit of Sec.~\ref{sec:strip} in 30-digit arithmetic, because the weights are large and of alternating sign. It agrees with the integral form to between $10^{-11}\%$ and $10^{-8}\%$ inside the strip and differs from it by 1800\% to 340\,000\% outside. Each term of the fit adds its own poles $k_{0N}+\jj/a_n$ and roots, and we identify root~3 among them by following it continuously from the single exponential as the radii are moved to those of the fit.

\textit{Singularities of the closed form.} Equation~\eqref{eq:closed} is analytic in $k$ except at three kinds of points. It has simple poles at $k=\pm\kappa_N$ for each channel, branch points at the same points through $\artanh(k/\kappa_N)$, and branch points at $k=\pm\jj/a$ through $\arctan(ka)$. The terms in $1/k^2$ to $1/k^5$ of $F_1$ and $F_2$ are each singular at $k=0$, but their singularities cancel. A branch point is the end of a branch cut, which is a line across which a multivalued function jumps from one of its values to another. The correct values follow from how the coherent field is obtained. The field radiated by a plane source is the inverse Fourier transform of $g_M=1/D_M$ along the real $k$ axis. When this path is moved upward into the complex plane, the field becomes a sum of residues at the roots of $D_M$, which are the coherent waves, plus integrals around the branch cuts. The functions must therefore take the values reached continuously from the real axis along vertical paths, with the branch cuts running vertically upward from $\kappa_N$ and $\jj/a$ (and downward from $-\kappa_N$ and $-\jj/a$), as shown in Fig.~\ref{fig:complex}(b). Near the real axis, these are the principal values. The principal value of $\arctan(ka)$ already has vertical branch cuts and applies everywhere, whereas $\artanh(k/\kappa_N)=\frac12[\mathrm{Ln}(\kappa_N+k)-\mathrm{Ln}(\kappa_N-k)]$, with $\mathrm{Ln}(x)=\ln(-\jj x)+\jj\pi/2$, the logarithm whose branch cut lies along the negative imaginary axis. The integrals around the branch cuts add a part of the coherent field that is not a sum of waves, which we do not evaluate in this paper.

\textit{Root count.} The number of roots follows from the argument principle, i.e.\ from the change in the phase of $D_M$ along a closed contour, divided by $2\pi$. The contour encloses the region $0<\Real k\,a<1.5\,k_{0\mathrm{T}}a+6$ and $0<\Imag k\,a<5$, and it excludes the branch cuts by narrow slits closed below each pole by a small semicircle. The closed form has no poles in this region, so the result is the number of roots. It is exactly three at every frequency examined, from $2k_{0M}a=0.1$ to 100, for transverse and longitudinal waves in iron, and no further roots appear when the region is extended to $\Imag k\,a<20$. In practice, we find the roots by Newton iteration at a low frequency, $2k_{0M}a=0.05$, starting from $k_{0M}$ for root~1 and from the poles $\kappa_N$, corrected to first order with Eq.~\eqref{eq:weak}, for roots~2 and 3, and we then follow each root in frequency.

\section{Meaning of the geometric pair}\label{app:pair}

When the incident wave scatters mainly into its own channel, roots~1 and 2 of the geometric regime have a simple meaning. Keeping only that channel, Eq.~\eqref{eq:geomcubic} becomes $(V_{0M}^2-V^2)^2=4Q_{M\to M}V_{0M}^4$, whose two roots are
\begin{equation}\label{eq:geompair}
  V=V_{0M}\sqrt{1\mp2\sqrt{Q_{M\to M}}}.
\end{equation}
For transverse waves in iron, Eq.~\eqref{eq:geompair} gives 2758 and 3641~m/s, within 0.7\% of roots~1 and 2 of Eq.~\eqref{eq:geomcubic}. These two velocities represent the spread of the wave speed from grain to grain, as we now show. With this channel only, Eq.~\eqref{eq:geomM} gives
\begin{equation}\label{eq:pairm}
  m_{M\to M}=\frac{k^4V_{0M}^4\sigma^2}{\omega^2-k^2V_{0M}^2},\qquad\sigma^2=4Q_{M\to M}.
\end{equation}
Equation~\eqref{eq:pairm} is identical to the mass operator of a scalar wave in a medium whose stiffness $\rho V_{0M}^2(1+\epsilon)$ varies from grain to grain, with grains much larger than the wavelength and a relative fluctuation $\epsilon$ of zero mean and standard deviation $\sigma$, the scalar case studied by Calvet and Margerin~\cite{Calvet2012}. The mean Green's function of Eq.~\eqref{eq:dispersion} then splits exactly into two terms,
\begin{equation}\label{eq:pairg}
  g_M=\frac{1}{2}\left[\frac{1}{\omega^2-k^2V_{0M}^2(1+\sigma)}+\frac{1}{\omega^2-k^2V_{0M}^2(1-\sigma)}\right].
\end{equation}
Each term is the Green's function of a homogeneous medium with $\epsilon=\pm\sigma$, i.e.\ with the velocities $V_{0M}\sqrt{1\pm\sigma}$ of Eq.~\eqref{eq:geompair}, so $g_M$ is the average over two equally likely values of $\epsilon$. Physically, a grain much larger than the wavelength carries the wave at its own local speed, so the exact mean Green's function would be the average over the actual distribution of $\epsilon$. The FOSA replaces this distribution by two values, one standard deviation on either side of the mean. The two velocities thus represent the spread of the local wave speed, whose relative standard deviation is, to first order, $\sigma/2=\sqrt{Q_{M\to M}}$. This is the inhomogeneity parameter of Calvet and Margerin~\cite{Calvet2012}.

\bibliography{references}

\end{document}